\documentclass[showpacs,10pt,twocolumn,prb]{revtex4-1}

\usepackage{amsmath}
\usepackage{amssymb}
\usepackage{graphicx}
\usepackage{amssymb}
\usepackage{graphics}
\usepackage{epsfig}
\usepackage{color}

\begin{document}

\title{Spin glass in semiconducting KFe$_{1.05}$Ag$_{0.88}$Te$_{2}$ single crystals}
\author{Hyejin Ryu,$^{1,2,\dag}$ Hechang Lei,$^{1,\S}$ B. Klobes,$^{3}$ J. B. Warren,$^{4}$, R. P. Hermann,$^{3,5,\ddag}$ and C. Petrovic$^{1,2}$}
\affiliation{$^{1}$Condensed Matter Physics and Materials Science Department, Brookhaven
National Laboratory, Upton, New York 11973, USA\\
$^{2}$Department of Physics and Astronomy, Stony Brook University, Stony Brook, New York 11794-3800, USA\\
$^{3}$J\"{u}lich Centre for Neutron Science JCNS and Peter Gr\"{u}nberg Institute PGI, JARA-FIT, Forschungszentrum J\"{u}lich GmbH, D-52425 J\"{u}lich, Germany\\
$^{4}$Instrument Division, Brookhaven National Laboratory, Upton, New York 11973, USA\\
$^{5}$Facult\'{e} des Sciences, Universit\'{e} de Li\`{e}ge, B-4000 Li\`{e}ge, Belgium}

\date{\today}

\begin{abstract}

We report discovery of KFe$_{1.05}$Ag$_{0.88}$Te$_{2}$ single crystals with semiconducting spin glass ground state. Composition and structure analysis suggest nearly stoichiometric I4/mmm space group but allow for the existence of vacancies, absent in long range semiconducting antiferromagnet KFe$_{0.85}$Ag$_{1.15}$Te$_{2}$. The subtle change in stoichometry in Fe/Ag sublattice changes magnetic ground state but not conductivity, giving further insight into the semiconducting gap mechanism.
\end{abstract}

\maketitle

\section{Introduction}

Since the discovery of the high temperature superconductor LaFeAsO$_{1-x}$F$_{x}$, superconductivity has been found in many iron pnictides with different crystal structures such as AFeAs (A = alkaline or alkaline-earth metal), and (AFe$_{2}$As$_{2}$, A= Ca, Sr, Ba, and Eu).\cite{Kamihara,Rotter,Wang} Iron chalcogenide materials, however, feature superconducting critical temperatures of up to about 30 K in bulk at high [FeCh (Ch = S, Se, and Te)] or ambient pressure [A$_{x}$Fe$_{2-y}$Se$_{2}$ (A = K, Rb, Cs, and Tl)] and over 100 K in thin films.\cite{Hsu,Medvedev,Guo,WangAF,Krzton-Maziopa,WangHD,GeJF} Among the most notable characteristics of iron chalcogenide superconductors are chemical inhomogeneity and deviations from ideal stoichiometry with considerable influence in magnetic interactions and superconductivity. Binary FeCh materials feature interstitial iron whereas ternary materials show vacancy-induced nanoscale separation on magnetic and superconducting domains.\cite{Bao0,Bao,Ryan,WangZ,LiW,DingX}

The existence of super-lattice of Fe-vacancies in (Tl,K,Rb)Fe$_{x}$Se${_2}$ system results in an occurrence of the block antiferromagnetic and semiconducting state.\cite{FangMH} Recently, it has been found that K$_{x}$Fe$_{2-y}$S$_{2}$ and KFe$_{0.85}$Ag$_{1.15}$Te$_{2}$ feature spin glass and long range magnetic order, respectively.\cite{Lei0,Lei1} The latter material, in particular, is a K or Fe/Ag vacancy-free and its magnetism and mechanism of non-metallic state is of high interest. Ag atoms fill Fe lattice so that there are no vacancies on Fe/Ag site in the crystal structure. Yet, Ag does mimic Fe vacancy in the electronic structure since Ag orbitals are sunk from the Fermi level. Thus Fe$^{2+}$ unconventional magnetic and insulating states can be studied in materials crystallizing in the Fe vacancy-free I4/mmm space group, identical to the space group of superconducting nano and micro -scale domains in A$_{x}$Fe$_{2-y}$Se$_{2}$.\cite{LiW,DingX,Ang,Texier,Lazarevic}

In this work we report discovery of semiconducting spin glass KFe$_{1.05}$Ag$_{0.88}$Te$_{2}$ single crystals with spin freezing temperature T$_{f}$ below $\sim$53 K in 1000 Oe. The material crystallizes in I4/mmm space group with possible vacancies on the metal site, demonstrating that magnetic ground state is very sensitive to the subtle ratio of Fe/Ag and defects.

\section{Experiment}

Single crystals of KFe$_{1.05}$Ag$_{0.88}$Te$_{2}$ were synthesized from nominal composition KFe$_{1.25}$Ag$_{0.75}$Te$_{2}$ as described previously.\cite{Lei1} Single crystals with typical size 2 $\times$ 2 $\times$ 0.5 mm$^{3}$ were grown. Powder X-ray diffraction (XRD) spectra were taken with Cu $K_{\alpha}$ radiation ($\lambda = 0.15418$ nm) by a Rigaku Miniflex X-ray diffractometer. The lattice parameters were obtained by refining XRD spectra using the Rietica software.\cite{Hunter} The element analysis was performed using an energy-dispersive X-ray spectroscopy (EDX) in JOEL LSM-6500 scanning electron microscope. Room temperature $^{57}$Fe M\"{o}ssbauer spectra were measured on a constant-acceleration spectrometer using a rhodium matrix $^{57}$Co source. The spectrometer was calibrated at 295 K with a 10 $\mu$m $\alpha$-Fe foil and isomer shifts are reported relative to $\alpha$-Fe. Magnetization measurements, electrical transport, and heat capacity were carried out in Quantum Design MPMS-XL5 and PPMS-9. The in-plane resistivity $\rho(T)$ was measured by a four-probe configuration on cleaved rectangular shape single crystals.

\section{Results and Discussion}

\begin{figure}
\centerline{\includegraphics[scale=0.32]{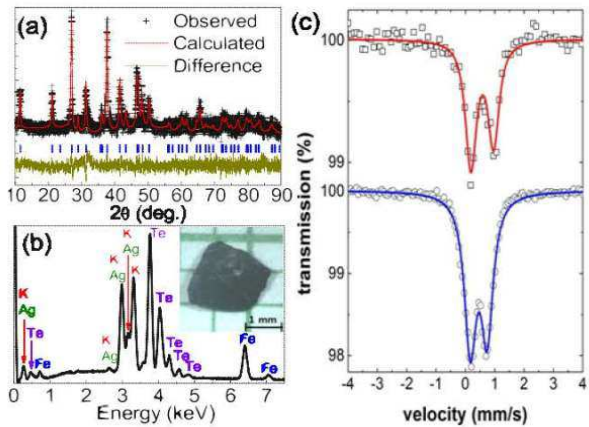}} \vspace*{-0.3cm}
\caption{(Color online) (a) Powder XRD patterns of KFe$_{1.05}$Ag$_{0.88}$Te$_{2}$. (b) The EDX spectrum of KFe$_{1.05}$Ag$_{0.88}$Te$_{2}$. The inset shows a photo of typical single crystal. (c) M\"{o}ssbauer spectrum of KFe$_{1.05}$Ag$_{0.88}$Te$_{2}$ (open squares) and KFe$_{0.85}$Ag$_{1.15}$Te$_{2}$ (open circles) at room temperature.}
\end{figure}

The refinement of crystallographic unit cell of KFe$_{1.05}$Ag$_{0.88}$Te$_{2}$ can be fully explained by I4/mmm space group [Fig. 1(a)]. The refined lattice parameters are a = 4.336(2) ${\AA}$ and c = 15.019(2) ${\AA}$. The value of a axis parameter is smaller while c axis lattice parameter is larger when compared to KFe$_{0.85}$Ag$_{1.15}$Te$_{2}$ [a = 4.371(2) ${\AA}$ and c = 14.954(2) ${\AA}$].\cite{Lei0} Also, they are smaller than the lattice parameter of CsFe$_{x}$Ag$_{2-x}$Te$_{2}$,\cite{Li} while larger than those of K$_{x}$Fe$_{2-y}$Se$_{2}$ and K$_{x}$Fe$_{2-y}$S$_{2}$,\cite{Guo,Lei1} since ionic size of K$^{+}$ is smaller than that of Cs$^{+}$, and ionic sizes of Ag$^{+}$ and Te$^{2-}$ are larger than ionic sizes of Fe$^{2+}$ and Se$^{2-}$(S$^{2-}$). EDX spectrum of single crystals shown in Fig. 1(b) confirms the existence of K, Fe, Ag, and Te. The average stoichiometry determined by EDX for several single crystals with multiple measuring points indicates that the crystals are homogeneous with K:Fe:Ag:Te=1.03(3):1.05(4):0.88(5):2.00 stoichiometry when fixing Te to be 2. The stoichiometry on Fe/Ag site is 1.93  (9) which suggests full occupancy but still allows for small deviations (vacancies) in contrast to KFe$_{0.85}$Ag$_{1.15}$Te$_{2}$.\cite{Lei0}

Room temperature M\"{o}ssbauer spectra (see table I for spectral parameters) of both KFe$_{1.05}$Ag$_{0.88}$Te$_{2}$ and KFe$_{0.85}$Ag$_{1.15}$Te$_{2}$ exhibit a doublet [Fig. 1(c)]. The unequal line intensities are due to preferred grain orientation in the powderized samples, as verified by a measurement with different angle between sample and incident beam direction.

\begin{table}[tbp]\centering%
\caption{Isomer shift $\delta$, quadruple splitting $\Delta E_Q$, and linewidth $\Gamma$ for KFe$_{1.05}$Ag$_{0.88}$Te$_{2}$ and KFe$_{0.85}$Ag$_{1.15}$Te$_{2}$.}%
\begin{tabular}{ccccc}
\hline\hline
  & $\delta$ (mm/s) & $\Delta E_{Q}$ (mm/s) & $\Gamma$ (mm/s)\\
\hline
KFe$_{1.05}$Ag$_{0.88}$Te$_{2}$  & 0.57(1) & 0.77(1) & 0.41(2) \\
KFe$_{0.85}$Ag$_{1.15}$Te$_{2}$ & 0.45(1) & 0.57(1) & 0.48(1) \\
\hline\hline
\end{tabular}%
\label{4}%
\end{table}%

Isomer shifts are slightly higher than those reported for other (metallic) ThCr$_{2}$Si$_{2}$ type compounds,\cite{Ryan,Nowik} but still confirm the divalent nature of Fe in these cases as no secondary Fe species could be detected. Moreover, comparable values for isomer shift and quadrupole splitting were reported for related compounds with mixed occupation of the Fe site.\cite{Nowik2} The latter aspect also manifests in the significantly increased linewidths. Although hyperfine parameters in Fe containing ThCr$_{2}$Si$_{2}$ compounds may strongly scatter,\cite{Ryan,Nowik,Nowik2,Nowik3} an increase of quadrupole splitting was also observed for K$_{0.8}$Fe$_{1.75}$Se$_{2}$ as compared to vacancy-free KFe$_{2}$Se$_{2}$\cite{Nowik3} and thus may support the assumption of vacancies in the KFe$_{1.05}$Ag$_{0.88}$Te$_{2}$ compound.

\begin{figure}
\centerline{\includegraphics[scale=0.9]{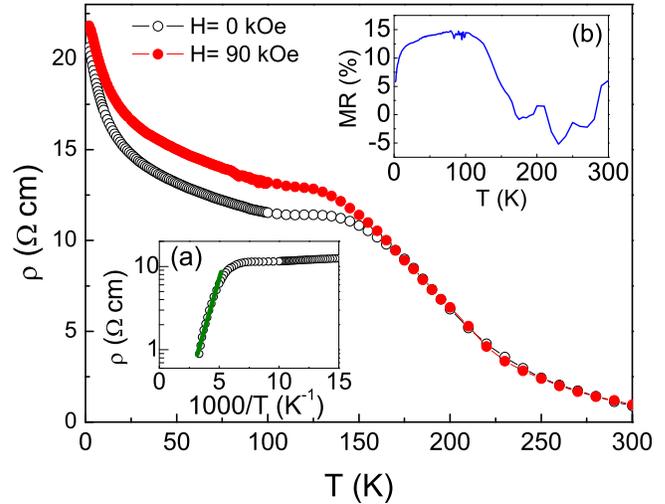}} \vspace*{-0.3cm}
\caption{(Color online) Temperature dependence of the in-plane resistivity of KFe$_{1.05}$Ag$_{0.88}$Te$_{2}$ with H = 0 kOe (open black circle) and 90 kOe (closed red circle) for H$\|$c direction. Inset (a) exhibits thermal activation model fitting (green solid line) for $\rho_{ab}(T)$ at H= 0 kOe. Inset (b) shows temperature dependence of magnetoresistance.}
\end{figure}

Temperature dependence of the in-plane resistivity of KFe$_{1.05}$Ag$_{0.88}$Te$_{2}$ single crystal is shown in Fig. 2. As temperature decreases, $\rho(T)$ increases with a shoulder appearing around 140 K. This is at somewhat higher temperature compared to KFe$_{0.85}$Ag$_{1.15}$Te$_{2}$.\cite{Lei0} The in-plane room temperature resistivity $\rho(T)$ is around 1 $\Omega cm$, similar to KFe$_{0.85}$Ag$_{1.15}$Te$_{2}$.\cite{Lei0} The $\rho(T)$ above 200 K can be fitted by thermal activation model $\rho=\rho_{0}exp(E_{a}/k_{B}T)$, where $\rho_{0}$ is a prefactor, $E_{a}$ is an activation energy, and $k_{B}$ is Boltzmann's constant (Fig.2 inset a). The obtained value of $\rho_{0}$ is 0.19(2) $\Omega cm$. This is larger than the value found in K$_{x}$Fe$_{2-y}$Se$_{2}$ and K$_{x}$Fe$_{2-y}$S$_{2}$. The gap value is $E_{a}$ = 43(2) $meV$, is smaller than the values in K$_{x}$Fe$_{2-y}$Se$_{2}$ and K$_{x}$Fe$_{2-y}$S$_{2}$.\cite{Lei0,Lei1} KFe$_{1.05}$Ag$_{0.88}$Te$_{2}$ single crystal shows pronounced magnetoresistance (MR) (Fig. 2 inset b) especially below 140 K similar to KFe$_{0.85}$Ag$_{1.15}$Te$_{2}$.\cite{Lei0} But unlike in KFe$_{0.85}$Ag$_{1.15}$Te$_{2}$, MR is positive suggesting weakened antiferromagnetic interactions in spin glass crystal.

\begin{figure}
\centerline{\includegraphics[scale=0.8]{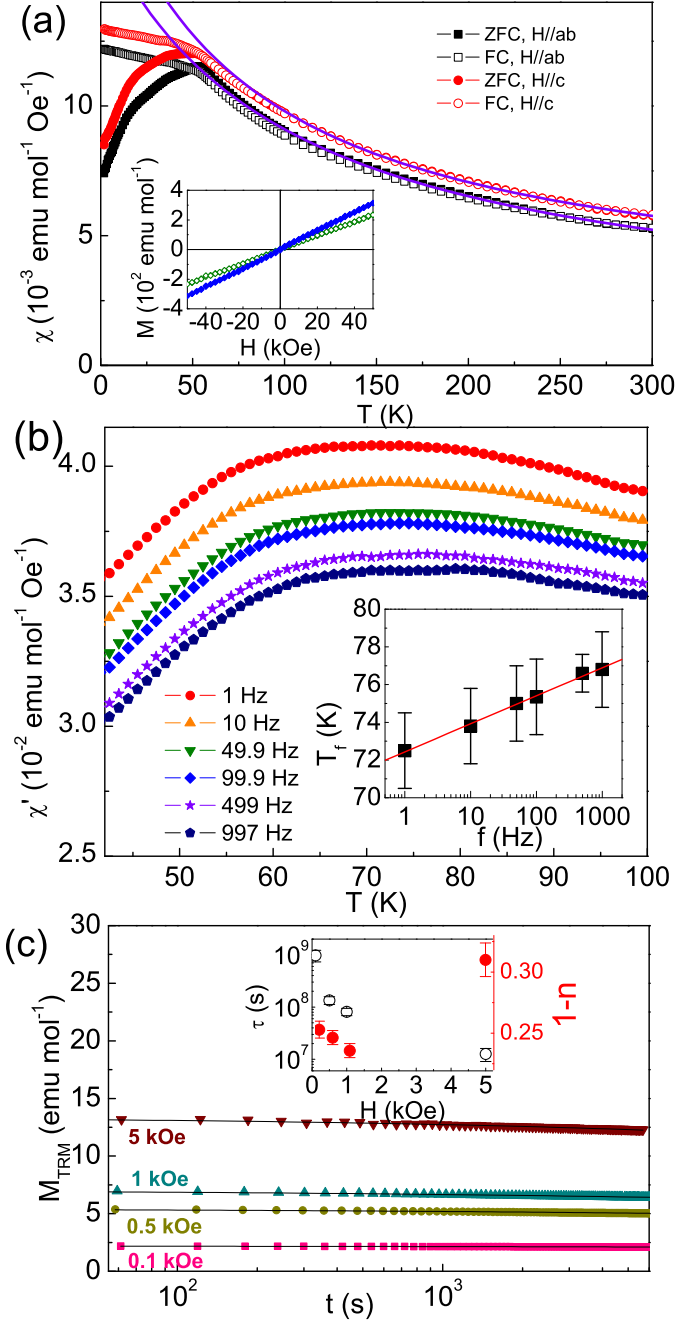}} \vspace*{-0.3cm}
\caption{(Color online) Magnetic properties of KFe$_{1.05}$Ag$_{0.88}$Te$_{2}$ single crystals. (a) Zero field cooled (ZFC) and field cooled (FC) anisotropic magnetic susceptibilities. The solid lines are Curie-Weiss fits. Inset shows M-H loops for H$\|$ab at 1.8 K (filled diamond) and 300 K (open diamond). (b) Temperature dependence of $\chi'(T)$ measured at several fixed frequencies taken in 3.8 Oe ac field. Inset is the frequency dependence of $T_{f}$ with the linear fit (solid line). The midpoint and temperature interval over which the $\chi'(T)$ takes its highest value were taken for $T_{f}$ and its error bar respectively. (c) Thermoremanent magnetization (TRM) at 10 K and $t_{w}=100s$ with different dc field and fits (solid lines). Inset is H-field dependence $\tau(s)$ (open circles) and 1-$n$ (filled circles).}
\end{figure}

The dc magnetic susceptibility of KFe$_{1.05}$Ag$_{0.88}$Te$_{2}$ for H$\|$c is slightly larger than H$\|$ab as shown in Fig. 3 (a). Both curves follow Curie-Weiss temperature dependence $\chi(T)=\chi_{0}+C/(T-\theta)$, where $\chi_{0}$ includes core diamagnetism, van Vleck and Pauli paramagnetism, $C$ is the Curie constant, and $\theta$ is the Curie-Weiss temperature. The obtained values are $\chi_{0}$ = 1.4(2)$\times$ 10$^{-3}$ emu mol$^{-1}$ Oe$^{-1}$, $C$ = 1.55(9) emu mol$^{-1}$ Oe$^{-1}$ K, and $\theta$ = -100(9) K for H$\|$ab, and $\chi_{0}$ = 2.1(1)$\times$ 10$^{-3}$ emu mol$^{-1}$ Oe$^{-1}$, $C$ = 1.38(7) emu mol$^{-1}$ Oe$^{-1}$ K, and $\theta$ = -80(7) K for H$\|$c. The effective moments obtained from the above values are $\mu_{eff}=1.57(2)\mu_{B}$/Fe for H$\|$ab and $\mu_{eff}=1.50(4)\mu_{B}$/Fe for H$\|$c. These are are smaller than expected for free Fe$^{2+}$ ions, smaller than in K$_{1.00(3)}$Fe$_{0.85(2)}$Ag$_{1.15(2)}$Te$_{2.00(1)}$\cite{Lei0} and even smaller than in a 3d spin 1/2 paramagnet ($\mu$$_{eff}$=1.73$\mu$$_{B}$). The irreversible behavior of $\chi(T)$ below 53 K in 1000 Oe implies ferromagnetic contribution or glassy transition. Similar behavior has been reported for KFeCuS$_{2}$, KFe$_{2}$Se$_{2}$, TlFe$_{2-x}$Se$_{2}$ and KMnAgSe$_{2}$.\cite{Oledzka,Lei1,Ying,LaiX} The magnetization loop is linear at 300 K while slightly curved s-shape at 1.8 K, also indicates possible spin glass system.\cite{Ying}

As frequency increases, the peak of the real part of the ac magnetic susceptibility $\chi'(T)$ shifts to higher temperature while the magnitude of $\chi'(T)$ decreases, which is a typical behavior of a spin glass.\cite{Mydosh} The frequency dependence of peak position (T$_{f}$) shown on Fig. 3 (b) is fitted by K=$\Delta$T$_{f}$/(T$_{f}$$\Delta$log$f$), and the obtained K value is 0.0201(2). This is in agreement with the values (0.0045 $\leq$ K $\leq$ 0.08) for a canonical spin glass.\cite{Mydosh} Fig. 3 (c) shows thermoremanent magnetization (TRM). The sample was cooled down from 100 K (above T$_{f}$) to 10 K (below T$_{f}$) in different magnetic fields, and kept there for $t_{w}=100s$. Then, the magnetic field was turned off and the magnetization decay M$_{TRM}$(t) was measured. At T = 10 K, M$_{TRM}$(t) shows slow decay, so M$_{TRM}$(t) has non-zero values even after several hours. This is fitted using a stretched exponential function, M$_{TRM}$(t) = $M_{0}exp[-(t/\tau)^{1-n}]$, where $M_{0}$, $\tau$, and 1-$n$ are the glassy component, the relaxation time, and the critical exponent, respectively. The obtained $\tau$ decreases up to 1 kOe and increases suddenly at 5 kOe, whereas 1-$n$ value keeps decreasing as H increases (Fig. 3 (c) inset) The attained 1-$n$ value is around 1/3, which is expected for a typical spin glass system.\cite{Campbell,ChuD} The spin glass behavior could arise from magnetic clusters due to Fe vacancies and disorder (similar to TlFe$_{2-x}$Se$_{2}$ when x$\geq$0.3 and KFe$_{2}$S$_{2}$)\cite{Ying,Lei1} or due to random distribution of magnetic exchange interactions on the metal sublattice as in KMnAgSe$_{2}$.\cite{LaiX}

\begin{figure}
\centerline{\includegraphics[scale=0.9]{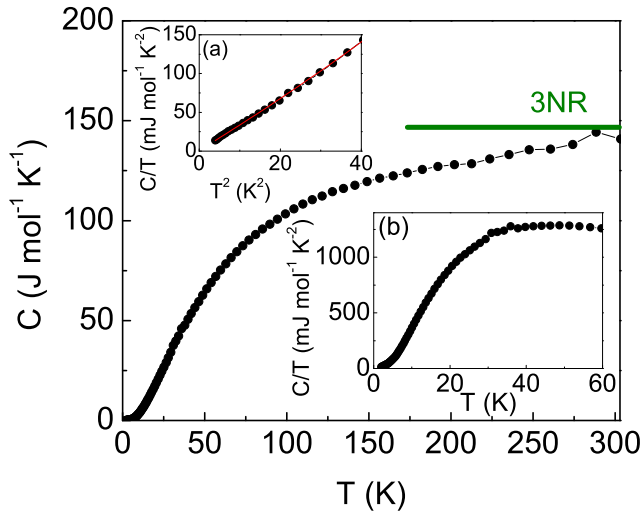}} \vspace*{-0.3cm}
\caption{(Color online) Temperature dependence of specific heat for KFe$_{1.05}$Ag$_{0.88}$Te$_{2}$ single crystal. Inset (a) shows the relation between C/T and T$^{2}$ at low temperature. The solid line represent fits by the equation C/T=$\gamma_{SG}$+$\beta$T$^{2}$. Inset (b) shows C/T vs. T relation at low temperature.}
\end{figure}

Heat capacity measured from T = 1.9 K to T = 300 K in zero magnetic field approaches the Dulong-Petit value of 3NR = 150 (J/mol K) at high temperatures (Fig. 4), where N is the atomic number and R is the gas constant. Low temperature heat capacity is fitted by C/T=$\gamma_{SG}$+$\beta$T$^{2}$ [Fig. 4 (a) inset] yielding $\gamma_{SG}$ = 0.88(6) mJ mol$^{-1}$ K$^{-2}$ and $\beta$ = 3.20(5) mJ mol$^{-1}$ K$^{-4}$. The Debye temperature can be estimated by $\Theta_{D}=(12\pi^{4}NR/5\beta)^{1/3}$ = 144.9(5) K. This is almost the same as in KFe$_{0.85}$Ag$_{1.15}$Te$_{2}$ single crystal and much smaller than $\Theta_{D}$ of K$_{x}$Fe$_{2-y}$Se$_{2}$ and K$_{x}$Fe$_{2-y}$S$_{2}$ possibly due to the larger atomic mass of Ag and Te. The nonzero value of $\gamma_{SG}$ is commonly found in magnetic insulating spin glass materials due to constant density of states of the low-temperature magnetic excitations.\cite{Meschede,Raju,Brodale}

When compared to KFe$_{0.85}$Ag$_{1.15}$Te$_{2}$, KFe$_{1.05}$Ag$_{0.88}$Te$_{2}$ shows more than twice larger values of room temperature resistivity, most likely due to possible  additional, vacancy induced disorder in the Fe/Ag sublattice occupation.\cite{Lei0} On the other hand the estimate of the energy gap size is larger in crystal with antiferromagnetic long range order. Optimal interlayer magnetic interaction plays a critical role in the appearance of the spin glass in KMnAgSe$_{2}$\cite{LaiX}, hence similar is expected in KFe$_{1-x}$Ag$_{x}$Te$_{2}$. Indeed, in the spin glass crystal the unit cell is elongated along the c-axis whereas the Fe plane is contracted when compared to the sample with long range order. The contraction of Fe plane suggests stronger covalent bonding, leading to increased electron density at the Fe site. This could explain reduced paramagnetic moment of Fe and smaller values of semiconducting gap. We note that band structure calculations indicate that KFeAgTe$_{2}$ with reduced Ag content could be more metallic.\cite{Shain}

\section{Conclusion}

In summary, we report on the discovery of semiconducting spin glass KFe$_{1.05}$Ag$_{0.88}$Te$_{2}$ single crystals. Composition and structure analysis implies I4/mmm space group with possible vacancies on Fe site. This is in contrast to KFe$_{0.85}$Ag$_{1.15}$Te$_{2}$ single crystals with long range antiferromagnetic order. The mechanism of semiconducting gap that arises due to electronic correlations (Mott vs. Hund mechanism) in KFe$_{1-x-\delta}$Ag$_{x}$Te$_{2}$ (where $\delta$ is putative vacancy) is of considerable interest in iron superconductors as well as in other correlated electron materials.\cite{Ang,Haule,YinZP,Mravlje} Since the Hund gap is sensitive to magnetic structure rather than Hubbard repulsion U, it would be instructive to further investigate electronic correlations and magnetic structure in KFe$_{1-x}$Ag$_{x}$Te$_{2}$ materials with variable Fe/Ag ratio.

\begin{acknowledgements}

Work at Brookhaven is supported by the U.S. DOE under Contract No. DE-AC02- 98CH10886 and in part
by the Center for Emergent Superconductivity, an Energy Frontier Research Center funded by the U.S. DOE, Office for Basic Energy Science (K. W and C. P). R.P.H. acknowledges support from the Helmholtz Association for the Helmholtz-University Young Investigator Group "Lattice Dynamics in Emerging Functional Materials"

\end{acknowledgements}

\dag Present address: Advanced Light Source, E. O. Lawrence Berkeley National Laboratory, Berkeley, California 94720, USA
\S Present address: Department of Physics, Renmin University, Beijing 100872, China
\ddag Present address: Materials Science and Technology Division, Oak Ridge National Laboratory, Oak Ridge, Tennessee 37831, USA

\end{document}